\documentclass[aip,
rsi,
graphicx,
reprint,
amsmath,
amssymb,
floatfix,
twocolumn
]{revtex4-1}

\draft % marks overfull lines with a black rule on the right

\usepackage{graphicx}
\usepackage{siunitx}
\usepackage{amsmath}
\usepackage[symbol]{footmisc}

\usepackage{hyperref}% add hypertext capabilities

\usepackage[T1]{fontenc}

\hypersetup{
    colorlinks=true,       % false: boxed links; true: colored links
    linkcolor=blue,          % color of internal links (change box color with linkbordercolor)
    citecolor=blue,        % color of links to bibliography
    filecolor=blue,      % color of file links
    urlcolor=blue           % color of external links
}

\begin{document}

%\preprint{}

\title{Programmable System on Chip for controlling an atomic physics experiment} %Title of paper



\author{A. Sitaram}
\email[]{asitaram@umd.edu}
\author{G. K. Campbell}
\author{A. Restelli}
\email[]{arestell@umd.edu}
%\email[]{gcampbe1@umd.edu}
\affiliation{Joint Quantum Institute, University of Maryland and National Institute of Standards and Technology, College Park, Maryland 20742, USA}


\date{\today}

\begin{abstract}
Most atomic physics experiments are controlled by a digital pattern generator used to synchronize all equipment by providing triggers and clocks.
Recently, the availability of well-documented open-source development tools has lifted the barriers to using programmable systems on chip (PSoC), making them a convenient and versatile tool for synthesizing digital patterns.
Here, we take advantage of these advancements in the design of a versatile clock and pattern generator using a PSoC.
We present our design with the intent of highlighting the new possibilities that PSoCs have to offer in terms of flexibility.
We provide a robust hardware carrier and basic firmware implementation that can be expanded and modified for other uses.
\end{abstract}


\maketitle %\maketitle must follow title, authors, abstract and \pacs

\section{Introduction}\label{intro}

Laser-cooled atoms, ions, and molecules are interesting and dynamic systems to study, and are being used to develop many quantum technologies.
These technologies include precise atomic clocks~\cite{Bothwell2019, Hinkley2013}, quantum computers and simulators~\cite{Cirac2012,Verstraete2009}, and quantum sensors~\cite{Ye2008,Sorrentino2009}.
Experiments in atomic, molecular, and optical (AMO) physics are often a combination of a large number of commercial or custom-made instruments from different sources and manufacturers that need to operate synchronously and in a repeatable fashion. 
Synchronization is achieved by using a specialized software suite to control a primary digital pattern generator or clock device with deterministic timing that sends trigger signals to the other hardware devices.
The PulseBlaster by SpinCore\cite{disclamer}, a commercial device based on a field programmable gate array (FPGA), is commonly used as a primary clock in many AMO experiments~\cite{Starkey2013} and is compatible with many different software suites.
Many university groups have also designed custom-made devices based around a microcontroller or an FPGA as their primary clock.
Microcontrollers combine processing power with many peripherals for interfacing directly with hardware, and have found use in a wide variety of physics experiments~\cite{Patel2020,Hosak2018,Gaskell2009,Malek2019}.
%with a relative ease of programming.
% Microcontrollers have found use in the control of electron spectrometer experiments in conjunction with LabView~\cite{Patel2020} and as an arbitrary digital pulse sequence generator with low jitter~\cite{Hosak2018}.
On the other hand, FPGAs provide versatility in modifying the overall system architecture to accommodate changes in functionality, although they require more expertise for development.
Despite the steeper learning curve, FPGAs have become a common choice as a control device in many physics experiments and work extremely well to accommodate more complex architectures, as well as modular ones~\cite{Bertoldi2020, Perego2018, Donnellan2019, Keshet2013, Lee2013}.

Another approach for controlling experiments is to create a complete infrastructure of software and modular hardware that is designed with built-in timing synchronization.
Two commercial examples of this approach are LabView, a systems engineering software that is compatible with National Instruments hardware, and ARTIQ by M-labs~\cite{Kulik2018}, which is also a complete infrastructure of software and hardware.
Some university research groups have also created complete architectures, basing their hardware designs off of FPGAs and designing custom control software~\cite{Perego2018, Keshet2013}.

While FPGAs can work well as a primary control device for an experiment, microcontrollers offer a simpler solution for handling complex communications protocols such as USB (Universal Serial Bus) or Ethernet.
Often, a microcontroller is used in conjunction with an FPGA, either externally\cite{Starkey2013} or instantiated within the FPGA\cite{Kulik2018}.
An alternative approach is to use a programmable system on chip (PSoC), which combines an FPGA and a high performance microprocessor on a single chip.
This allows implementation of operating systems, advanced communication protocols, and high level language interpreters in the microprocessor, leveraging the FPGA when hardware acceleration or control of dedicated peripherals is needed.
% PSoCs have been incorporated by some groups in conjunction with the Experimental Physics and Industrial Control System (EPICS) software~\cite{Xue2015, Lee2016}. 
Previously, development using PSoCs has been less accessible due to the baseline level of expertise required, but recently, thanks to the diffusion and level of maturity of tools such as PetaLinux~\cite{Petalinux} or Yocto Project~\cite{Yocto} for the generation of GNU/Linux images, PSoCs have become more widely adopted~\cite{Xue2015, Lee2016}.
%{Previously, development using PSoCs has been less accessible due to the baseline level of expertise required, but recently, they have become more widely used and have been incorporated by some groups in conjunction with the Experimental Physics and Industrial Control System (EPICS) software~\cite{Xue2015, Lee2016}.}
% \begin{itemize}
%   \item {There is need of $\mu$C for communications and other functionalities and an external $\mu$C is used in most of projects with FPGA[ref where it is the case]}
%   \item {it is also possible to add a soft uC in the FPGA, for example Artiq is based on that [cite]}
%   \item {The other option is the increasilgly popular PSoC and this has been used by [cite EPIC]}
%   \item {Costs has reduced for PSoC}
%   \item {the advantage is that with respect an external uC the PSoC is connected with the programmable logic more tightly allowing for a higher communication bandwidth}
%   \item {performance compared with a softcore is higher and consistent because there is no dependency oh how the core is placed and routed into the FPGA}
%   \item {availability of open source solutions to run operating systems on the processor such as GNU/Linux and run applications and interpreted languages. Petalinux, Yocto Linux (https://www.yoctoproject.org/), PYNQ, etc...}
%   \item{ we based our project a PSoC... etc....}
% \end{itemize}

We chose a PSoC architecture to design our 64-channel pattern generator and primary clock with the goal of expanding the capabilities of our ultracold strontium experiment.
Our requirement, to have a large number of channels operating in parallel with fast (\SI{100}{\nano\second} resolution) and deterministic timing, points towards an FPGA as the platform of choice; however, we also had the goal of handling most of the data communication protocols using high level abstraction languages, such as Python, to facilitate testing and future rapid development.
To achieve these goals, we take advantage of the PYNQ (Python Productivity for Zynq)  infrastructure~\cite{PYNQ}, a platform for the development of applications with the Xilinx Zynq series of programmable systems on chip based on GNU/Linux and Python.
Our lab uses the Labscript Suite of software~\cite{Starkey2013} to control our experiment, which uses a text and GUI approach to provide efficient experimental control for atomic physics experiments and is based on the Python programming language.
We designed a hardware platform around a Microzed Zynq-7020 module~\cite{Microzed} (produced by Avnet) mounted on a custom carrier board with four low-jitter input trigger lines and eight breakout boards with eight channels each to route the 64 output lines.
The FPGA gateware is written in Verilog and System Verilog, and we used Xilinx native development tools in order to make use of the many verification features, such as complex testbenches for behavioral simulation.
In the next sections, we will describe the system architecture as a whole, as well as describe the hardware and firmware in detail.

\section{System Overview}\label{overview}

\begin{figure}[h!]
  \includegraphics[width = \columnwidth]{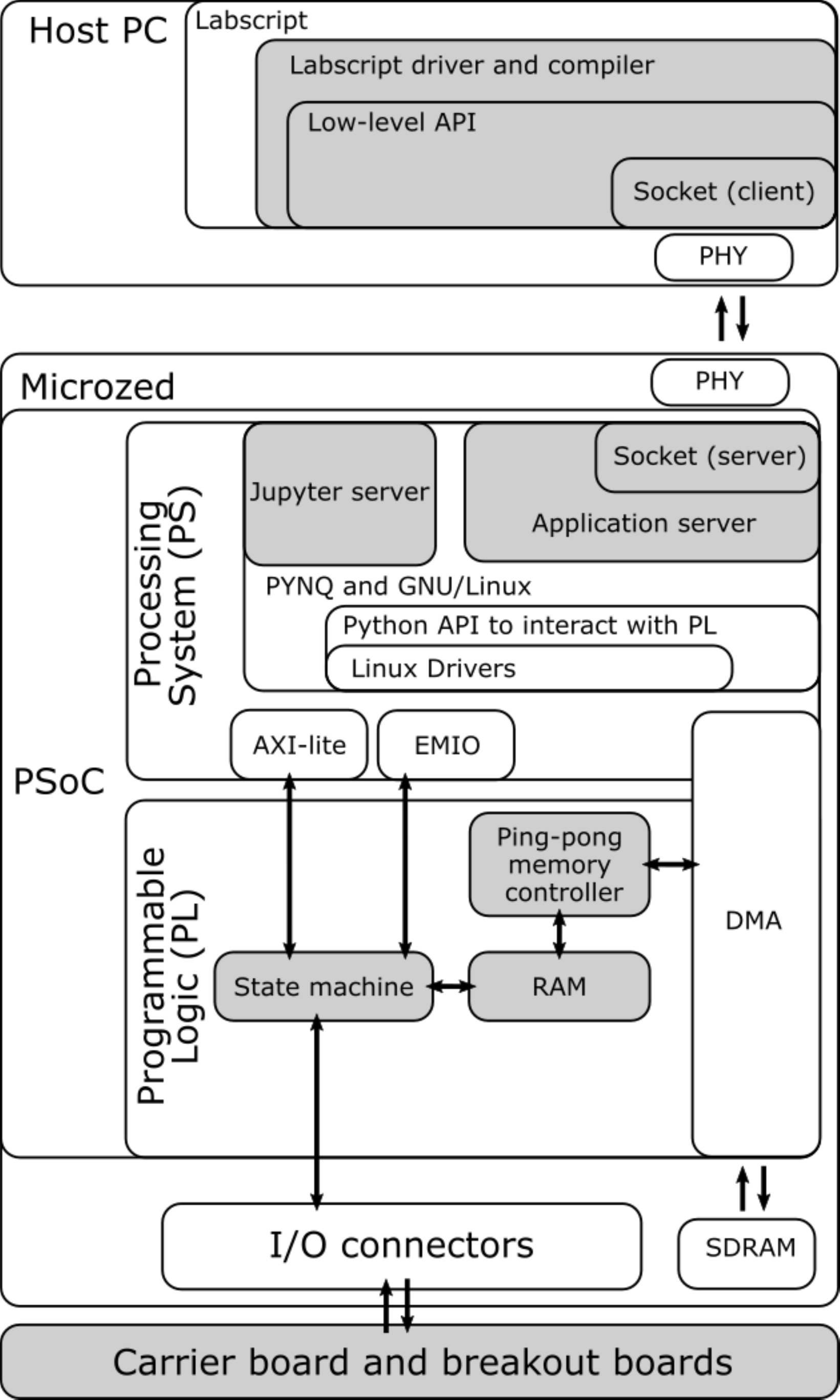}
  \caption{\label{fig:schematic} 
  Overall schematic of the pattern generator. For convenience, we summarize acronyms used in the figure: PC (Personal Computer), API (Application Programming Interface), PHY (PhYsical interface), PSoC (Programmable System on Chip), PS (Processing System), PL (Programmable Logic), AXI (Advanced eXtensible Interface), EMIO (Extended Multiplexed Input/Output), DMA (Direct Memory Access), RAM (Random Access Memory), I/O (Input/Output), SDRAM (Synchronous Dynamic RAM).
  Elements of the system we designed in detail are shown in gray, while the white blocks are the components and software that are available as commercial modules, open-source libraries, or automatic software generation tools.
  }
\end{figure}

Fig.~\ref{fig:schematic} illustrates the overall architecture of our design, which can be broken down into three blocks: our host PC (or lab control computer), the Microzed-7020 module, and the carrier and breakout boards.
The Microzed module contains the Xilinx PSoC and a series of additional peripherals, of which we show only the most relevant to our project: the I/O connectors to interface with the carrier board, 1 GB of synchronous dynamic random access memory (SDRAM), and an Ethernet physical layer chip (PHY) used for communication with the host PC.
The PSoC (Xilinx XC7Z020-1CLG400C~\cite{XC7020}) is composed of the processing system (PS) and the programmable logic (PL).
The PS is a dual core ARM Cortex-A9, while the PL is an Artix-7 FPGA fabric with approximately 85000 logic cells.      

The PYNQ ecosystem allows us to run Linux Ubuntu on the PS and is equipped with a Jupyter notebook server accessible from a remote machine browser as a means to interact with the PL using a Python application programming interface (API). 
Through the API, the PL can be accessed using extended multiplexed input/output lines (EMIO) or an AXI-lite (Advanced eXtensible Interface) channel that can be used to map configuration registers in the PL to the operating system's RAM.
Additionally, the SDRAM external memory used by the operating system can be accessed using a direct memory access (DMA) controller.

The heart of our design is in the PL, where we implemented a state machine written in System Verilog which reads instructions from RAM instantiated in the FPGA fabric.
To allow for a longer list of instructions, we have implemented a ping-pong memory controller that moves data from the external SDRAM to the PL RAM through the DMA channel.
The state machine and ping-pong memory controller will be discussed in further detail in Sections~\ref{statemachine} and~\ref{pingpong}, respectively.

In the PS, we wrote an application server in Python to receive instructions from the host PC through a socket connection, transferring them to the shared SDRAM and initiating DMA transfers.
The application server is paired with a socket client running on the host PC, also written in Python, which acts as a low-level API to interface the Labscript instrument driver with the Microzed module. 

The Microzed board plugs into a custom-designed carrier board using MicroHeader connectors.
The output signals are then routed through eight breakout boards and are accessible via BNC (Bayonet Neill–Concelman) connectors. The carrier board and breakout board designs are described in Sec.~\ref{hardware}.

\section{Hardware Features}\label{hardware}

\subsection{Carrier Board}\label{carrier}
\begin{figure}[!thb]
  %\begin{flushleft}(a) \hspace{1.5in} (b)\end{flushleft}
  \includegraphics[width = \columnwidth]{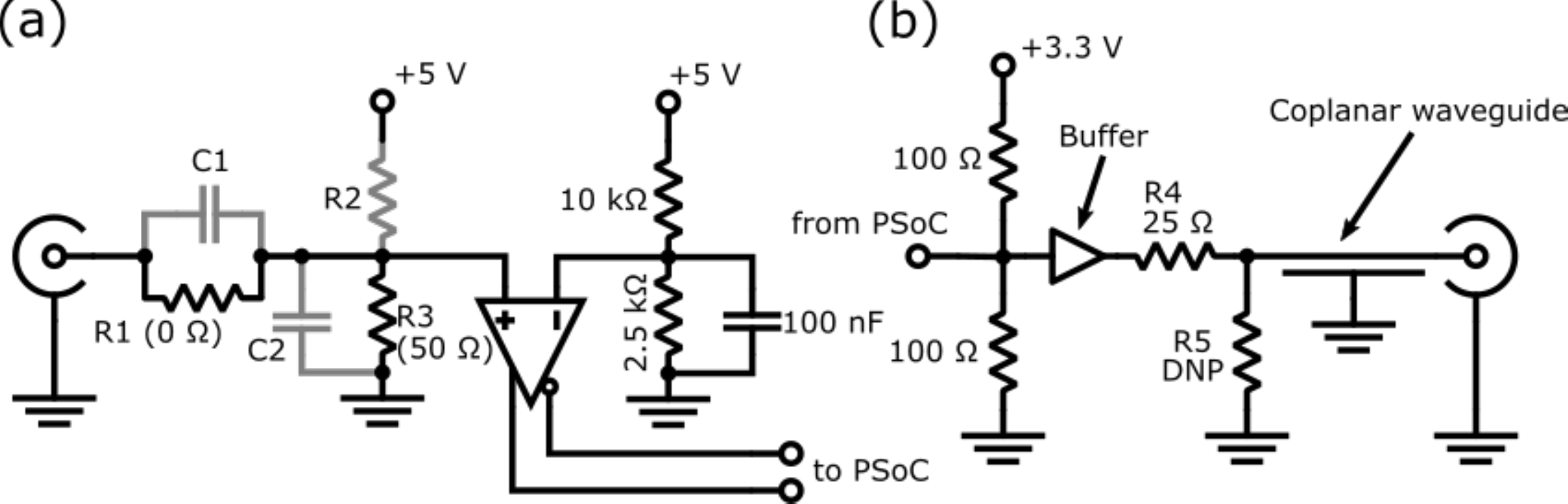}
  
  \caption{\label{fig:terminations} 
  Termination networks used to interface the FPGA logics with external signals. (a) shows the circuit used for the four digital inputs on the carrier board while (b) shows the circuit used for the 64 digital outputs. }
\end{figure}

The carrier PCB (Printed Circuit Board) routes the 64 digital output lines from the expansion connectors of the Microzed module (Amphenol ICC 61083-101400LF) to eight 20-pin rectangular connectors, which are used to distribute the signals to the breakout boards using ribbon cables.
Placement of the 20-pin rectangular connectors was determined to keep the difference in length between all traces below \SI{12}{\milli\meter}.
The resulting maximum difference in propagation time between channels is estimated to be only $\approx$~\SI{64}{ps}, which is well within the goals of our design.  
The carrier board also has four BNC connectors for introducing input clock or trigger signals to the Microzed.
In order to adapt arbitrary trigger and clock standards to the FPGA input standards, each BNC input is connected to the analog front end circuit shown in Fig.~\ref{fig:terminations}(a).  
Input signals are sent through a high speed comparator chip (ADCMP552BRQZ~\cite{ADCMP552BRQZ}) with PECL (Positive Emitter-Coupled Logic) outputs. 
We set a \SI{1}{\volt} threshold on the inverting input of the comparator using a voltage divider filtered with a \SI{0.1}{\micro\farad} capacitor, and we connect the coaxial input to a network of components (R1, R2, R3, C1, C2) that can be used to adapt a variety of AC (Alternating Current) or DC (Direct Current) input waveforms. 
R1 is used as jumper to select between DC and AC inputs.
In the default DC-coupled configuration, R1~=~\SI{0}{\ohm} and R3~=~\SI{50}{\ohm}, making the input compatible with \SI{3.3}{\volt} and \SI{5}{\volt} TTL (Transistor Transistor Logic) standards.
For an AC-coupled configuration, R1 is not placed, C1~=~\SI{0.1}{\micro\farad} to block DC signals, and the values, R2~=~\SI{294}{\ohm} and R3~=~\SI{60.4}{\ohm}, set the input impedance to \SI{50}{\ohm}, maintaining an average voltage of \SI{0.85}{\volt} at the input of the comparator. 
The carrier board also provides a \SI{3.3}{\volt} supply for the I/O banks of the PSoC with two high-efficiency micro DC-DC converters (XCL214\cite{torex}), and a supervisor chip (STM6779LWB6F\cite{supervisor}) ensures that the required power sequencing for the PSoC is respected\cite{sequencing}.

\subsection{Breakout Boards}\label{breakout}

\begin{figure}[b]
  \includegraphics[width = \columnwidth]{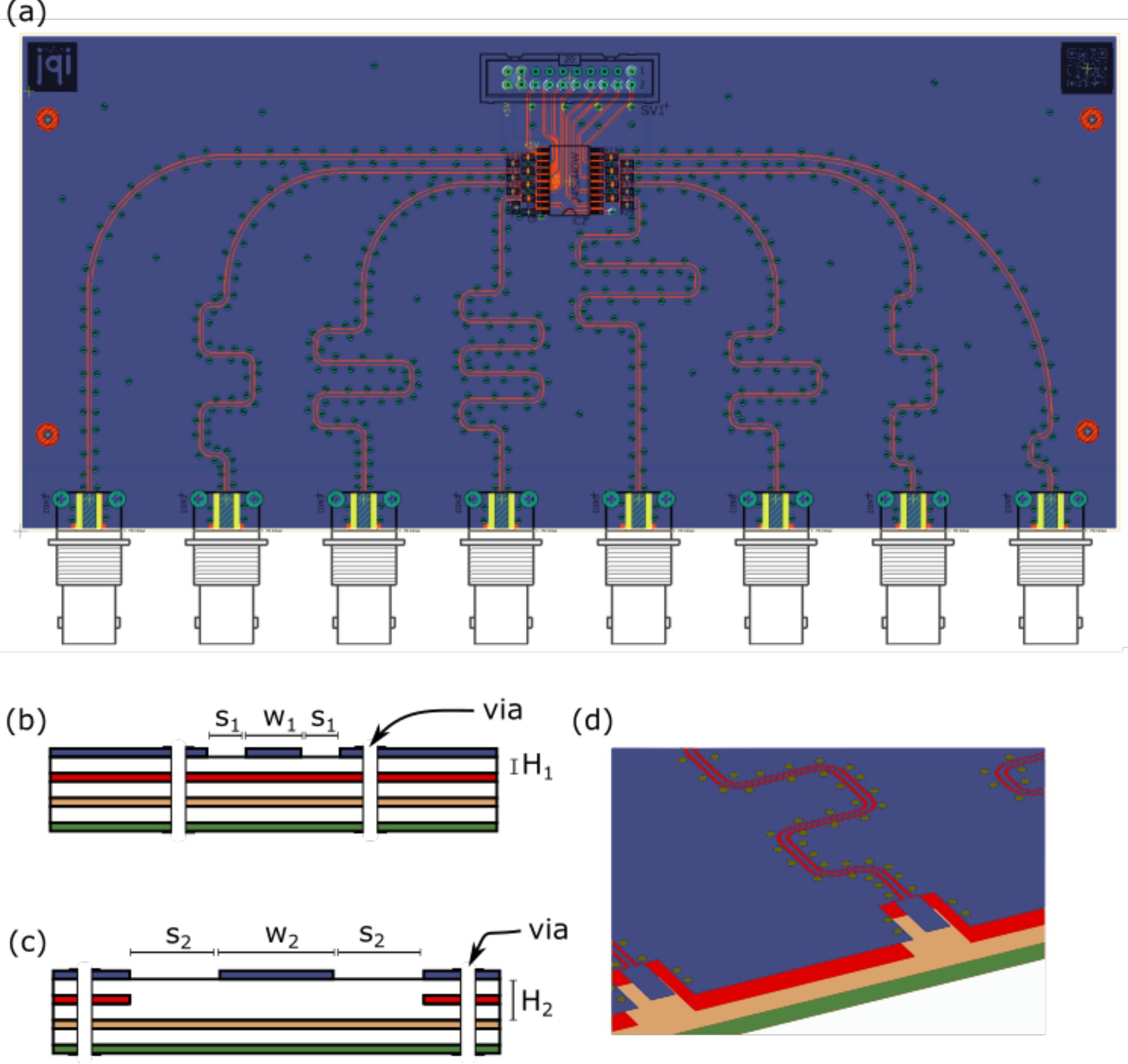}
  \caption{\label{fig:Breakout}
  (a) Breakout board layout. 
  Signals enter the board through the 20-pin connector at the top.
  Meanders help equalize electrical delays of all traces. 
  (b) Stack-up of the PCB for the coplanar waveguide
  (c) Stack-up of the PCB for the microstrip below the BNC connector.
  (d) Close-up perspective view of the circuit board layout.
  Figure is not to scale.}
\end{figure}

The eight breakout boards use a Texas Instruments octal buffer (SN74S244DWG4\cite{SN74S244DWG4}) to drive \SI{5}{\volt} TTL signals through \SI{50}{\ohm} coaxial cables. 
The electrical schematic for a single channel is shown in Fig.~\ref{fig:terminations}(b).
% Of particular importance was the design of the impedance control and shielding of the digital lines to minimize reflections and crosstalk. 
The ribbon cable connecting the carrier board with the breakout board has an alternating pattern of GND lines and digital signal lines, which prevents crosstalk and sets a characteristic impedance of \SI{50}{\ohm}. 
The ribbon cable also carries a \SI{3.3}{\volt} supply used for termination and a \SI{5}{\volt} supply used to power the octal buffer. 
The two \SI{100}{\ohm} resistors in Fig.~\ref{fig:terminations}(b) terminate  the single-ended line from the PSoC to a Thevenin equivalent of \SI{50}{\ohm} at half the logic supply. 
This type of termination is called split termination and is described on page 26 of the Xilinx UG471 user guide~\cite{UG471}.
Each output of the octal buffer has an internal impedance of \SI{25}{\ohm}, and therefore a series resistance of \SI{25}{\ohm} (R4) is added in order to bring the output impedance to a standard value of \SI{50}{\ohm}. 
The additional DNP (do not place) resistor (R5) can be used in conjunction with a different value for R4 to produce an arbitrary Thevenin equivalent output that maintains a \SI{50}{\ohm} impedance, allowing the user to configure the outputs to different logic standards.
For example, the values R4~=~\SI{75}{\ohm} and R5~=~\SI{100}{\ohm} would reduce the output voltage by a factor of 2. 
The coplanar waveguide in Fig.~\ref{fig:terminations}(b) is designed with a target impedance of \SI{50}{\ohm} using the Kicad PCB calculator~\cite{Kicad} software. 

\begin{figure}[t]
  \includegraphics[width = \columnwidth]{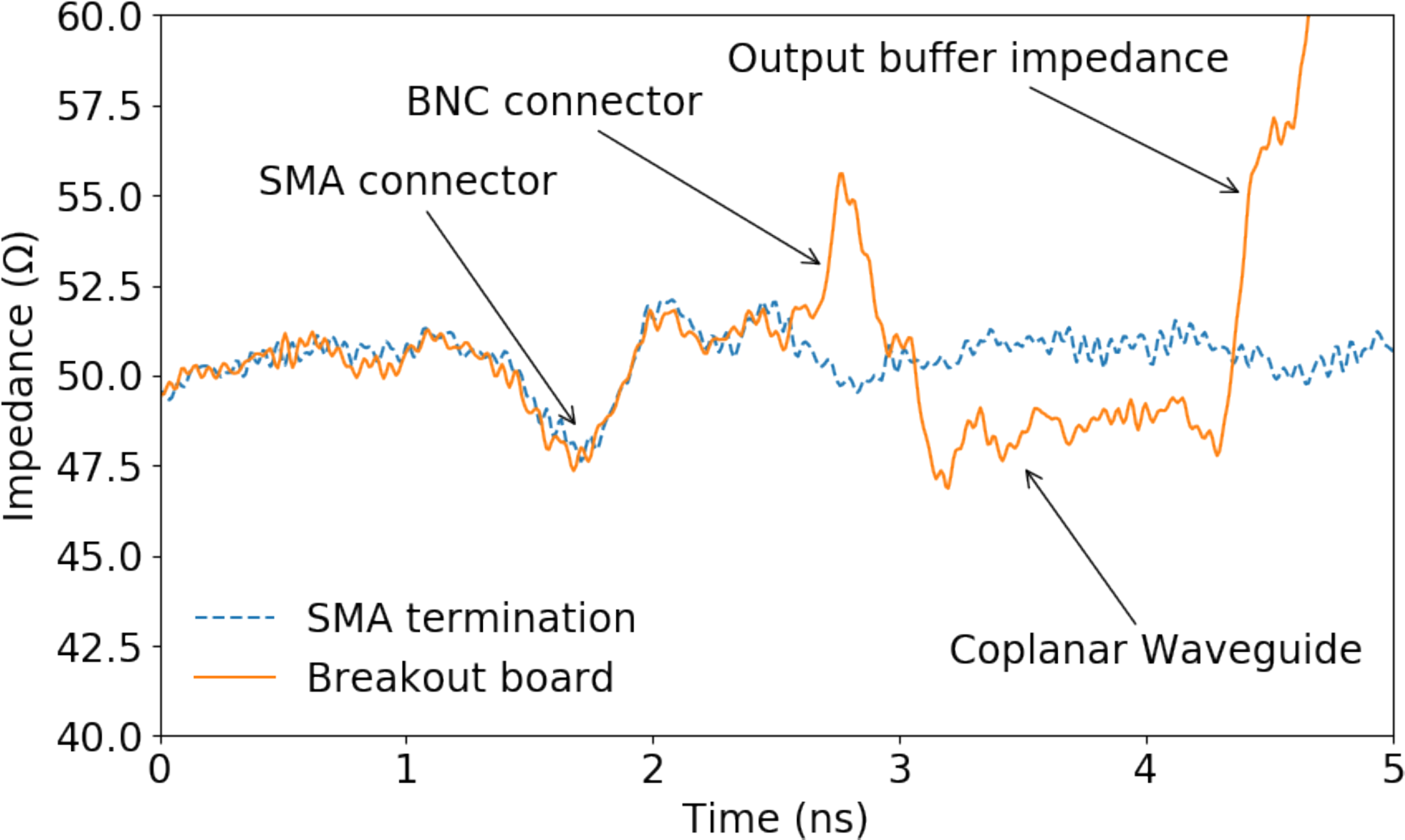}
  \caption{\label{fig:TDR} 
  Time Domain Reflectometry measurement of the breakout board.
  The characteristic impedance of the BNC connector and coplanar waveguide remain within $10\%$ of the \SI{50}{\ohm} target value.}
\end{figure}

Fig.~\ref{fig:Breakout}(a) shows how the eight coplanar waveguides are arranged on the breakout board. 
To prevent variations in the timing delay across different output channels, we have matched the length of all 8 traces using meanders. 
Based on the information provided by the PCB manufacturer (nominal relative dielectric constant $\epsilon_r$~=~4.3), we designed the coplanar waveguide, as illustrated in Fig.~\ref{fig:Breakout}(b), with a width $\text{W}_1$~=~\SI{0.46}{\milli\meter}, spacing between traces and top-layer ground plane $\text{S}_1$~=~\SI{0.3}{\milli\meter}, and separation from top-inner-layer ground plane $\text{H}_1$~=~\SI{0.24}{\milli\meter}.
We chose edge-mount BNC connectors rated up to 4 GHz to minimize the characteristic impedance discontinuity from the PCB to the coaxial cables. 
For impedance matching, the connectors need to be soldered to a microstrip that ends at the edge of the PCB.
However, to ensure an adequate mechanical strength for the connector's central pin soldering joint, the width of the microstrip must be much larger than the width $\text{W}_1$~=~\SI{0.46}{\milli\meter} of the coplanar waveguide in Fig.~\ref{fig:Breakout}(b). 
To allow for a wider section of the transmission line, we therefore remove the inner-top ground layer from under the central pin's soldering pad, as shown in Fig.~\ref{fig:Breakout}(c). 
Using the inner-bottom layer as the new ground plane, the distance from the transmission line is increased to $\text{H}_2$~=~\SI{1.26}{\milli\meter}. 
A nominal \SI{50}{\ohm} impedance is now obtained with $\text{W}_2$~=~\SI{2.29}{\milli\meter} and $\text{S}_2$~=~\SI{1.27}{\milli\meter}.
A perspective view of the PCB layers is shown in Fig.~\ref{fig:Breakout}(d).

We verified the performance of the transmission lines and BNC launch by performing a time domain reflectometry (TDR~\cite{HewletPackard}) measurement on the PCB.
The result of the meaurement is shown in Fig.~\ref{fig:TDR}, where we use the technique described in Ref.~\cite{HewletPackard} to measure the amplitude of a reflected step signal to calculate the characteristic impedance along a transmission line as a function of electrical delay.
We first measure the response of a coaxial cable with an SMA (SubMiniature version A) connector attached to a SMA~\SI{50}{\ohm} termination.
We then connect the coaxial cable to our PCB board, while not powered, using a SMA to BNC adapter and compare the two TDR responses. 
Four different sections can be distinguished in the traces in Fig.~\ref{fig:TDR}: the SMA connector, the BNC adapter, the coplanar waveguide on the PCB, and the output buffer passive impedance. 
Apart from the output buffer, which shows a change of impedance compatible with a capacitive load, the maximum impedance variation for the BNC connector and coplanar waveguide design is below $\approx 10\%$, limiting reflections below $\approx 5\%$.
% To verify the absence of crosstalk, we monitored one channel while actively changing all 63 of the other channels on an oscilloscope, and 

\section{Firmware Development}\label{firmware}

 \begin{figure*}[t]
   \includegraphics[width = \textwidth]{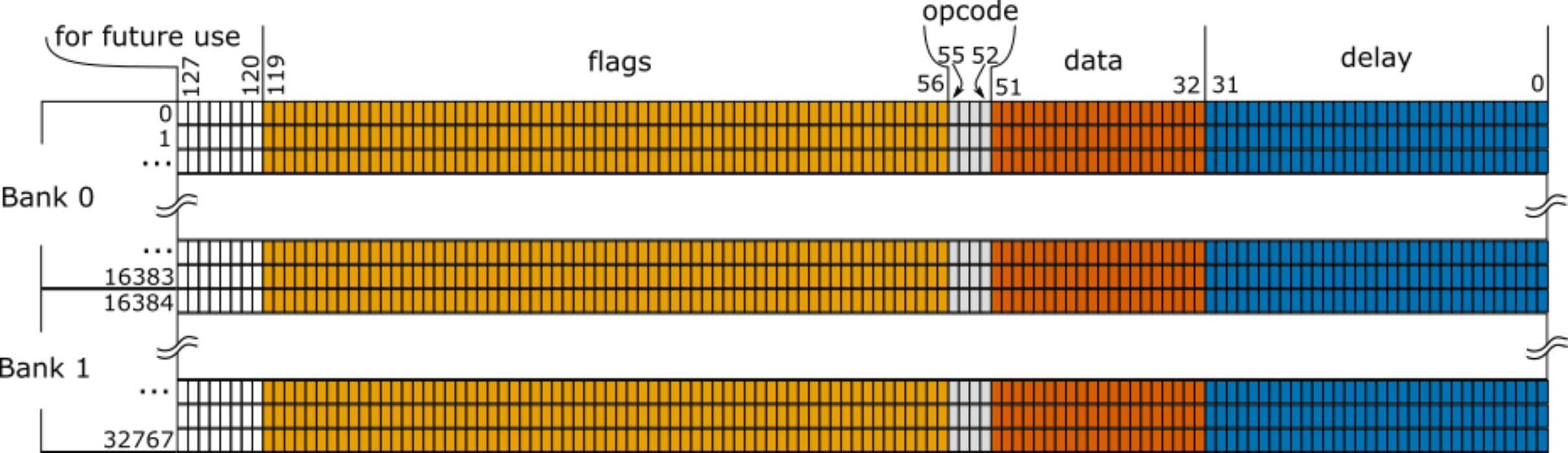}
   \caption{\label{fig:memmap} 
   Illustration of the memory in the FPGA. 
   The memory is split into Bank 0 and Bank 1, each with 16834 instructions.
   The memory has a width of 128 bits.
   Each instruction contains 64 bits for the state of each of the flags, 4 bits for the opcode, 20 bits for the data argument, and 32 bits for the time delay argument.
   The last 8 bits are left unused, but can be allocated in the future.}
 \end{figure*}

\subsection{Communication}\label{comm}

To communicate instructions to the PSoC, we open a socket server on the PS.
We then wait for the TCP/IP client on the lab computer to connect.
Once the connection is established, data is sent through the socket stored in a numpy array~\cite{NumpyArrays}, which is mapped on a contiguous section of SDRAM shared with the PL through a DMA controller.
The data is then accessed by the PL and processed by the state machine as instructions in a 128 bit format extension of the 80 bit long instruction format used in the PulseBlaster~\cite{PulseBlaster}. 
% that can be interpreted by the state machine.
In case the connection is unexpectedly broken, we have implemented an algorithm for the server to automatically refresh the same socket connection, instead of creating a new one.
This makes the system robust against the interruption of the connection without having to manually reset it.

\subsection{State Machine}\label{statemachine}

\begin{table}[b]
%\scriptsize
\centering\footnotesize
\begin{tabular}{ |p{0.75cm}|p{1.65cm}|>{\raggedright\arraybackslash}p{2.35cm}|>{\raggedright\arraybackslash}p{3cm}|}

\hline
\textbf{State}& \textbf{Instruction} & \textbf{Data} & \textbf{Function}\\
\hline
0 & CONTINUE & None & Continues to next instruction \\
\hline
1 & STOP & None & Stops execution of program \\
\hline
2 & LOOP & Number of desired loops, great than or equal to 1 & Specifies beginning of loop \\
\hline
3 & END LOOP & Address of beginning of loop & Specifies end of loop \\
\hline
4 & JSR & Address of first subroutine instruction \hfill \hfill & Jumps to a subroutine \\
\hline
5 & RTS & None & Program execution returns to instruction after JSR was called at the end of subroutine   \\
\hline
6 & BRANCH & Address in memory to branch to & Program execution branches to an address specified by data \\
\hline
7 & LONG \newline DELAY & Delay multiplier & Executes the length of instruction given in the time field multiplied by delay multiplier\\
\hline
8 & WAIT & None & Waits for a hardware trigger to continue program execution\\
\hline
\end{tabular}
\small
\caption{\label{tab:opcodes}
List of states that was programmed in the state machine with associated data field and description of the function performed.
The state numbering corresponds to the associated opcode.
%List of instructions and associated op codes, data, and function that was programmed into the PL of the FPGA.
  }
\end{table}

To control the 64 TTL output channels, we have written a Mealy~\cite{Mealy1955} state machine in the programmable logic of the FPGA.
In contrast with Moore~\cite{Moore56} state machines, Mealy state machines' inputs directly affect the outputs, allowing for a lower-latency design.
We wrote our state machine in System Verilog to take advantage of specialized features of the language, such as enumeration logic and the passing of structured data through design modules.
The state machine first fetches 128 bit instructions from a 128x32768 RAM, mapped as shown in Fig.~\ref{fig:memmap}.
There are five fields that make up the 128 bit instruction to the state machine: time delay (32 bits), data (20 bits) opcode (4 bits), flags (64 bits) and finally the remaining 8 bits are reserved for future use.
The state machine reads the memory bank row by row.
The opcode tells the state machine which state to enter next, and the flags field designates which output channels will be changed or affected with each instruction.
The data contains any special information specific to the current opcode.
For example, if the state machine is being instructed to enter a loop, the data would contain the number of loop iterations.
Finally, the `delay' argument indicates how long the state machine should wait before loading the next instruction.
The states that we have programmed in our state machine are shown in Table~\ref{tab:opcodes}, along with the accompanying `data' field.
To facilitate integration with Labscript, we choose an instruction set that is mostly compatible with the one of the Pulseblaster, which is extensively used within the Labscript codebase (we did not implement nested loops, as they are not used in Labscript).

\subsection{Ping-Pong Memory}\label{pingpong}

\begin{figure}[!ht]
  \includegraphics[width = \columnwidth]{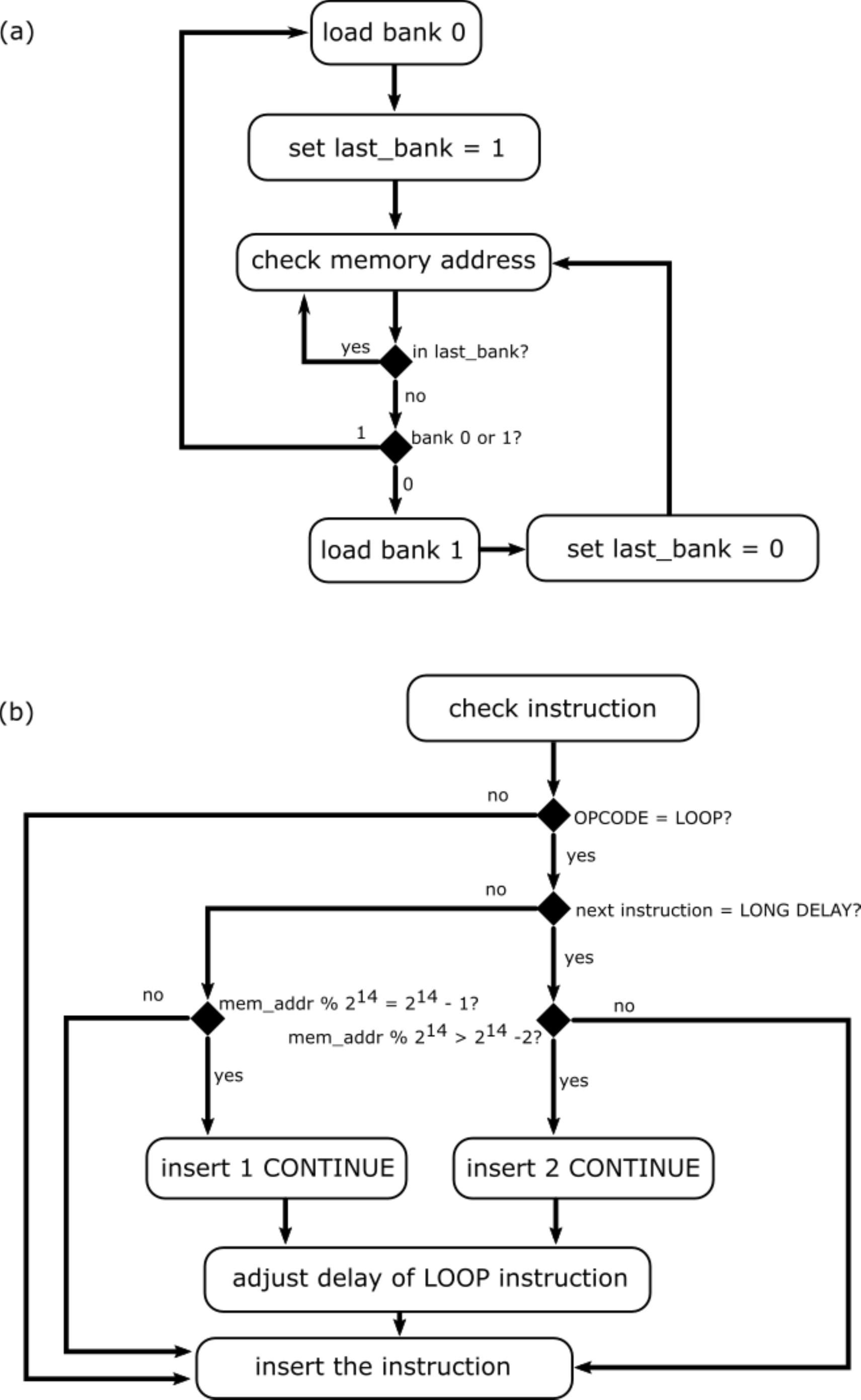}
  \caption{\label{fig:pingpong}
  Bank switching and compiler check algorithms for the ping-pong memory controller.
  (a) The system begins by loading bank zero and setting the ``last\_bank'' to Bank 1.
  From there, the system consistently checks the memory address of the state machine to determine whether it has switched banks in the memory.
  If it has switched banks, it changes ``last\_bank'' and loads the other bank of memory with new instructions.
  (b) Checks performed during compilation to avoid memory underflow. 
  The logic expressions mem\_addr~\%~$2^{14} > 2^{14} - 2$ and mem\_addr~\%~$2^{14} = 2^{14} - 1$ check, respectively, if mem\_addr is mapped to the last two slots or the last slot in the memory bank (\% is the \protect\begin{footnotesize}MOD\protect\end{footnotesize} operator).}
\end{figure}

The state machine described in the previous section is designed to read instructions from a 32768-instruction static memory.
To increase the available memory, we use the 32768-instruction space as a cache memory and divide it into two banks with $2^{14}$~=~16384 instructions each: Bank 0 and Bank 1, as shown in Fig.~\ref{fig:memmap}.
We then implement a ping-pong memory controller to automatically update the content of the memory by requesting direct memory access (DMA) to a large shared contiguous portion of the SDRAM, which has space for up to 8192000 instructions.
% ensures the available space greatly exceeds our system's needs.
The algorithm for the ping-pong memory controller is shown in Fig.~\ref{fig:pingpong}(a).
The controller begins by transferring 16384 instructions from SDRAM into Bank 0 of the PL RAM and setting a register called ``last\_bank'' equal to 1.
The main state machine then begins executing instructions from RAM, starting from Bank 0, while the ping-pong memory controller constantly monitors the memory address.
Each time the memory address is not in the bank identified by the register ``last\_bank'', the previously accessed bank is refreshed with new data from the SDRAM and the value of ``last\_bank'' is updated with the identifier of the currently accessed bank.
Setting ``last\_bank'' equal to 1 when the state machine starts causes Bank 1 to be updated immediately after Bank 0 as soon as the state machine accesses the memory.
The PL RAM is a dual port memory that can be independently addressed from two different clock domains. 
Thus, the state machine controlling the 64 TTL outputs does not need to be synchronous with the rest of the PL and with the PS. 
The ping-pong memory controller and DMA engine are clocked by the PS, while the state machine can be optionally clocked from one of the PLLs (Phase Locked Loop) available in the PL fabric that can be locked to an external reference connected to one of the four available BNC inputs. 

The automatic RAM refresh implemented by the ping-pong memory controller can pose a problem if certain instructions span over two banks, such as \begin{footnotesize}LOOP/END LOOP, BRANCH, JSR/RTS\end{footnotesize}.
For example, if a loop is started in the first bank, but ends in the second bank, since the first bank is updated with new instructions while the second bank is running, the system will no longer have the initial loop instruction to refer back to.
The compiler must be aware of this type of memory bank underflow or overflow and be able to resolve them by altering the order and number of instructions, without changing the final behavior at run time.
In the current Labscript driver, there is only one instance where underflow can happen: when the complex instruction called ``reps'' is translated into either a \begin{footnotesize}LOOP\end{footnotesize} immediately followed by an \begin{footnotesize}END LOOP\end{footnotesize} instruction or a series of \begin{footnotesize}LOOP, LONG DELAY, END LOOP\end{footnotesize}. 
To prevent memory underflow, we have implemented checks in the code while the program is compiling. 
The algorithm is illustrated in Fig.~\ref{fig:pingpong}(b).
When a \begin{footnotesize}LOOP\end{footnotesize} opcode is found, the system checks if either the instruction is mapped on the last instruction of a bank or if it is mapped on the second to last and is immediately followed by a \begin{footnotesize}LONG DELAY\end{footnotesize} instruction.
In these cases, it inserts additional \begin{footnotesize}CONTINUE\end{footnotesize} instructions to ensure that the \begin{footnotesize}LOOP\end{footnotesize} instruction is moved to the beginning of the next bank. 
To ensure that the insertion does not modify the original timing, the field `delay' in the \begin{footnotesize}LOOP\end{footnotesize} instruction is reduced by the duration of the inserted \begin{footnotesize}CONTINUE\end{footnotesize} instructions.

\section{Discussion}\label{discussion}

The PSoC-based primary clock device, that we have created for controlling AMO physics experiments, is easily integrated with the Labscript Suite.
The hardware provides 64 buffered digital outputs for controlling other hardware devices and also 4 input trigger channels.
The printed circuit board design ensures signal integrity and minimal crosstalk between channels.
Our firmware design implements a state machine written in System Verilog and a ping-pong memory controller that allows the execution of a large number of instructions (exceeding 8192000).
The system is currently being used to run the entire experiment in our lab, providing triggers for digital to analog coverters (DAC), digital direct synthesizers (DDS), mechanical shutters, and many other instruments.

According to the Synthesis tools timing reports the maximum frequency the state machine can operate at is \SI{104}{\mega\hertz}, and it is currently clocked at \SI{100}{\mega\hertz}.
Therefore, the current timing resolution is \SI{10}{\nano\second}, although using serializers in the PL fabric would allow timing resolutions down to \SI{1}{ns}.
The versatility of the platform also allows for other modifications, such as the possibility to add additional instructions to the state machine.
For example, an additional instruction could initiate a train of a specific number of pulses with an adjustable duty cycle and period using a single instruction, rather than using loops.
Other extensions of the instruction set could allow for conditional branching, which has already been shown to be useful in ion trapping experiments~\cite{Wright2017}. 
Further modifications to the design might include network security protocols and encryption for data transmission, which we have not included since our setup is running on an isolated network.
A possible use of the system we have considered, and have extensively taken advantage of during testing, is its capability to run scripts directly from the local Jupyter notebook server. 
With the Jupyter web interface, a remote computer is not necessary for the generation of patterns, and the device can be used as a stand-alone testbench digital pattern generator.

Our PSoC-based primary clock device has the capability to be integrated with many experimental setups with minimal modification, and the whole design is available online~\cite{ourdesign}.

\section*{Acknowledgements}

We thank Daniel Barker, Qi-Yu Liang, and Peter Elgee for their careful reading of the manuscript.
This work was partially supported by the NSF through the Physics Frontier Center at the Joint Quantum Institute.

\section*{Data Availability}

The data that support the findings of this study are available from the corresponding author upon reasonable request.

\bibliographystyle{aipnum4-1}
\bibliography{jane}

\begin{thebibliography}{39}
\bibitem [{dis()}]{disclamer}%
  \BibitemOpen
  \href@noop {} {}\bibinfo {howpublished} {Certain commercial products or
  company names are identified here to describe our study adequately. Such
  identification is not intended to imply recommendation or endorsement by the
  National Institute of Standards and Technology, nor is it intended to imply
  that the products or names identified are necessarily the best available for
  the purpose.}\BibitemShut {Stop}%
\end{thebibliography}

\end{document}